\documentclass[12pt,a4paper]{article}
\pdfoutput=1

\usepackage[T1]{fontenc}
\usepackage{lmodern}
\usepackage{amsmath}
\usepackage{amsfonts}
\usepackage{amssymb}
\usepackage{mathrsfs}
\usepackage{physics}
\usepackage{graphicx}
\usepackage{caption}
\usepackage{subcaption}
\usepackage{jheppub}
\usepackage{enumitem}
\usepackage{framed,float}
\usepackage{booktabs}
\usepackage{array}
\usepackage{soul}
\usepackage{multirow}
\usepackage{comment}
\usepackage{braket}
\usepackage{todonotes}
\usepackage{youngtab}
\usepackage{tikz}
\usetikzlibrary{calc,arrows,decorations.markings}

\usepackage{color}

\preprint{}
\title{\boldmath Duality between Seiberg-Witten Theory and Black Hole Superradiance}

\author[a]{Xian-Hui Ge,}
\author[b]{Masataka Matsumoto}
\author[a,c,d]{and Kilar Zhang}
\affiliation[a]{Department of Physics and Institute for Quantum Science and Technology, Shanghai University, 99 Shangda Road, Shanghai 200444, China}
\affiliation[b]{Wilczek Quantum Center, School of Physics and Astronomy, Shanghai Jiao Tong University, Shanghai 200240, China}
\affiliation[c]{Shanghai Key Lab for Astrophysics, Shanghai Normal University, 100 Guilin Road, Shanghai 200234, China}
\affiliation[d]{Shanghai Key Laboratory of High Temperature Superconductors, Shanghai 200444, China}
\emailAdd{gexh@shu.edu.cn}
\emailAdd{masataka@sjtu.edu.cn}
\emailAdd{kilar@shu.edu.cn}

\abstract{The newly established Seiberg-Witten (SW)/Quasinormal Modes (QNM) correspondence offers an efficient analytical approach to calculate the QNM frequencies, which was only available numerically before.  This is based on the fact that both sides are characterized by Heun-type equations.
 We find that a similar duality exists between Seiberg-Witten theory and black hole superradiance, since the latter can also be linked to confluent Heun equation after proper transformation. Then a dictionary is constructed, with the superradiance frequencies written in terms of gauge parameters.
Further by instanton counting, and taking care of the boundary conditions through connection formula, the relating frequencies are obtained analytically, which show consistency with known numerical results.}

\begin{document}

\newtheorem{theorem}{Theorem}

\newcommand{\gmf}[1]{\Gamma \left( #1 \right)}
\newcommand{\ba}{\begin{eqnarray}}
\newcommand{\ea}{\end{eqnarray}}
\newcommand{\nn}{\nonumber}
\newcommand{\cW}{{\mathcal{W}}}
\newcommand{\cO}{{\mathcal{O}}}
\newcommand{\cF}{{\mathcal{F}}}
\def\CY{\mathcal{Y}}
\def\CQ{\mathcal{Q}}
\def\CD{\mathcal{D}}

\newcommand{\be}{\begin{equation}}
\newcommand{\ee}{\end{equation}}
\newcommand{\hJ}{\hat{J}}
\newcommand{\hL}{\hat{L}}
\newcommand{\vY}{\vec{Y}}
\newcommand{\vW}{\vec{W}}
\newcommand{\bs}{\backslash}
\newcommand{\sqz}{\mathbf{z}}
\newcommand{\leg}{\mathrm{Leg}}
\newcommand{\arm}{\mathrm{Arm}}
\newcommand{\rrangle}{\rangle\!\rangle}
\newcommand{\llangle}{\langle\!\langle}

\newcommand{\cS}{\mathcal{S}}
\newcommand{\cD}{\mathcal{D}}
\newcommand{\cN}{\mathcal{N}}

\def\a{\alpha}
\def\b{\beta}
\def\c{\varepsilon}
\def\d{\delta}
\def\e{\epsilon}
\def\eone{\epsilon_{1}}
\def\etwo{\epsilon_{2}}
\def\f{\phi}
\def\g{\gamma}
\def\h{\theta}
\def\k{\kappa}
\def\l{\lambda}
\def\m{\mu}
\def\n{\nu}
\def\p{\psi}
\def\q{\partial}
\def\r{\rho}
\def\s{\sigma}
\def\t{\tau}
\def\u{\upsilon}
\def\v{\varphi}
\def\w{\omega}
\def\x{\xi}
\def\y{\eta}
\def\z{\zeta}
\def\D{\Delta}
\def\G{\Gamma}
\def\H{\Theta}
\def\L{\Lambda}
\def\F{\Phi}
\def\P{\Psi}
\def\S{\Sigma}
\def\o{\over}

\def\hg{\hat\gamma}

\def\Zv{\mathcal{Z}_\text{vect}}
\def\Zbf{\mathcal{Z}_\text{bif}}
\def\Zf{\mathcal{Z}_\text{fund}}
\def\bZbf{\bar{\mathcal{Z}}_\text{bif}}

\def\aY{|\vec{a},\vec{Y}\rangle}
\def\Ga{|G,\vec{a}\rangle}
\def\bGa{\langle G,\vec{a}|}
\def\CV{\mathcal{V}}
\def\tN{\tilde{N}}
\def\Zi{\mathcal{Z}_{\text{inst}}}
\def\nQ{Q}
\def\mf{m^{(f)}}

\def\CA{{\mathcal{A}}}
\def\CZ{{\mathcal{Z}}}
\def\CB{{\mathcal{B}}}
\def\CH{{\mathcal{H}}}
\def\CI{{\mathcal{I}}}

\newcommand{\no}{\nonumber}
\newcommand{\ov}[1]{{\overline{#1}}}
\newcommand{\superp}[2]{\genfrac{}{}{0pt}{}{#1}{#2}}

\def\la{\left\langle}
\def\ra{\right\rangle}

\newcommand{\EMA}{\mathsf{A}}                     
\newcommand{\EMF}{\mathsf{F}}                     

\newcommand{\ri}{\textrm{i}}

\allowdisplaybreaks
\maketitle
\flushbottom


\setcounter{footnote}{0}
\section{Introduction}

The recently developed Seiberg-Witten (SW) \cite{Seiberg:1994rs,Seiberg:1994aj}/Quasinormal Modes (QNM) \cite{vishveshwara1970scattering,Kokkotas:1999bd,Berti:2009kk,Konoplya:2011qq,Horowitz:1999jd}  correspondence \cite{Aminov:2020yma} provides a powerful tool for studying black holes (BH) and gravitational waves (GW) \cite{GW150914}. The binary black holes coalescences are the main sources of GW, and to interpret the observation data, we need first establish the waveform bank. There are three phases during the collision procedure: inspiral, merger and ringdown, and it is in the final stage, ringdown phase, that QNMs take over. While its spectra are known in high precision, those could only be achieved by numerical calculations.
However, since both the SW and QNM sides yield Heun type equations  \cite{Heun:1888}, by exactly matching the coefficients, one can derive the QNM frequencies analytically, with the help of instanton counting \cite{Nekrasov:2009rc} or thermodynamic Bethe ansatz (TBA) \cite{Gaiotto:2009hg} on the gauge theory side. This efficient method was quickly extended to other geometry besides the original Schwarzschild BH, Kerr BH and (A)dS BH cases, to Kerr-Newman BH, D3-branes, CCLP BH \cite{Bonelli:2021uvf,Bianchi:2021mft} and the charged
 C-metric \cite{Lei:2023mqx}, etc.. Besides, this duality offers a new way to verify the  gauge/CFT correspondence (AGT conjecture) \cite{Alday:2009aq}, and serve as an test ground for the connection formula \cite{Bonelli:2022ten} linking different boundary conditions.

Another hopeful way to find new physics through the SW/QNM correspondence is the study of BH superradiance \cite{Brito:2015oca, East:2017ovw}. Superradiance refers to a physical phenomenon that occurs when waves interact with rotating objects, such as Kerr BH. This interaction leads to the extraction of energy and angular momentum from the rotating object, which subsequently amplifies the scattered waves. In the realm of BH, superradiance is particularly intriguing as it offers a theoretical mechanism for extracting energy from the BH, potentially leading to evaporation in certain scenarios. The consideration of ultralight bosons \cite{Essig:2013lka} in proximity to a rotating BH further enhances the interest in superradiance. Ultralight bosons are hypothetical particles postulated to have masses significantly lighter than any particle currently known in the Standard Model of particle physics. If these particles exist and engage in interaction with a rotating BH, they undergo a process known as superradiant scattering, which results in the extraction of energy and angular momentum from the black hole. The investigation of superradiance via the SW/QNM correspondence  in this context holds the potential to yield profound insights into the fundamental nature of BH and their interplay with matter and radiation. Furthermore, it offers a promising avenue for indirectly detecting ultralight bosons through their interactions with BH. This area of theoretical physics research is currently active, and while direct experimental confirmations of ultralight bosons or superradiance effects around BH remain elusive, the possibility remains a highly intriguing prospect for future physics research.

Moreover, the BH superradiance is supposed to be probed by GW observation \cite{Baumann:2018vus}, adding an alternative target for the existing and future GW detectors \cite{Hu:2017mde,Ruan:2018tsw,TianQin:2015yph,LISA:2017pwj}. Superradiance will reduce the amplitude of the GW signal, and could also be revealed through finite-size effects like multipole moments and tidal Love numbers (TLN) \cite{Hinderer:2007mb}. Since in Einstein gravity this TLN vanishes for black holes \cite{Kol:2011vg,Gurlebeck:2015xpa}, a non-zero result offers an interpretation besides dark stars \cite{Zhang:2023hxd} for gap events like GW190814 \cite{LIGOScientific:2020zkf} or PSR J0514-4002E \cite{Barr:2024wwl}. Those analysis requires accurate calculation of frequencies to get precise waveforms. Currently, the method for obtaining the frequencies can be analytical done only for the near or far field limit, while in the intermediate range it is connected by boundary conditions, and finally solved by numerical process. Consequently, a full range analytic approach to superradiance frequencies will be a significant advance.

In this paper, we delve deeper into the intriguing duality between superradiance and SW/QNM correspondence.  This duality offers a new perspective on BH physics and supersymmetric gauge theory, revealing unexpected connections between these two seemingly disparate fields. Specifically, we demonstrate that under certain conditions, the scattering of waves from a rotating BH can lead to both superradiant amplification and the excitation of QNM. Furthermore, we show that this process is intimately related to the SW curve, which provides a geometric framework for understanding the underlying physics.  In section \ref{sec:2}, we explore the intricate relationship between the Seiberg-Witten curve and the confluent Heun equation, and summarize the existing knowledge on quasinormal modes of Kerr black holes. This section serves as a foundation for understanding the mathematical structure underlying the duality we explore, setting the stage in the context of the duality with superradiance. Section \ref{sec:3} and \ref{sec:4} are the cores of our paper. In section \ref{sec:3} we engage deeply in the study of superradiance through the SW/QNM correspondence. 
We unpack the details of the duality, revealing how superradiance, the SW curve, and quasinormal modes are intricately intertwined. In section \ref{sec:4} we carefully evaluate the boudary conditions with the help of connection formula, then show the results by instanton counting, and make comparison with known numerical studies.
In section \ref{sec:5}, we present our conclusions, summarizing the key findings and implications of our analysis. This section also highlights the significance of our work and points to future directions for further exploration. Necessary mathematical tools and derivations are collected in the appendix.

\section{Advances on Seiberg-Witten Curve and Quasinormal Modes} \label{sec:2}
In this section, we briefly review the duality between SW curve and BH quasinormal modes initiated in  \cite{Aminov:2020yma}.
\subsection{Seiberg-Witten curve and confluent Heun equation}
Seiberg-Witten theory is introduced to solve four dimensional $\mathcal{N}=2$ asymmetric gauge theory by evaluating the singularities and asymptotic behaviors. We leave the details to the references \cite{Seiberg:1994rs,Seiberg:1994aj}, and here focus on SU(2) theory with flavor number $N_f=3$, which is our main concern in this paper. The quantum SW curve in this case can be rewritten in the following form after some reparameterizations:

\begin{equation}
\begin{aligned}
\hbar^2 \psi''(z)+\left(\frac{1}{z^{ 2}(z-1)^2}\sum_{i=0}^4 \widehat{A}_i z^i \right)\psi(z)=0,
\end{aligned}
\end{equation}
which is in the normal form of the so-called confluent Heun equation \eqref{CHEnormal}  \cite{decarreau1978formes}.
The coefficients are expressed in gauge parameters by
\begin{equation}
\begin{aligned}
\widehat{A}_0&=-\frac{(m_1-m_2)^2}{4}+\frac{\hbar^2}{4},\\
\widehat{A}_1&=-E-m_1 m_2-\frac{m_3 \Lambda_3}{8}-\frac{\hbar^2}{4},\\
\widehat{A}_2&=E+\frac{3m_3 \Lambda_3}{8}-\frac{\Lambda_3^2}{64}+\frac{\hbar^2}{4},\\
\widehat{A}_3&=-\frac{m_3 \Lambda_3}{4}+\frac{\Lambda_3^2}{32},\\
\widehat{A}_4&=-\frac{\Lambda_3^2}{64} ,
\end{aligned}
\end{equation}
where $m_1$, $m_2$ and $m_3$ are the (anti-) fundamental hypermultiplets masses, while $\Lambda_3$ represents a function of the gauge coupling constant. $E$  stands for the eigenenergy, and $\hbar$ is the Plank constant.

On a torus the contour can be chosen as the so-called A-circle and B-circle, and the SW spectrum is characterized by the quantization condition
\be
\Pi_{I}^{(N_{f})}(E, {\bf m}, \Lambda_{N_f}, \hbar) =N_I  \left( n+\frac{1}{2} \right), \quad I =A,B, \quad n=0,1,2,\cdots,
\ee
with $N_A =i$ and $N_B =2\pi$ in our setup.

The quantum  A-period is 
\be
\label{atoF} 
\Pi_{A}^{(N_f)}(E,{\bf m}, \Lambda_{N_f},\hbar)=a(E,{\bf m},\Lambda_{N_f}, \hbar),
\ee
and the quantum B-period is given by
\be 
\label{BtoF} 
\Pi_{B}^{(N_f)}(E,{\bf m}, \Lambda_{N_f},\hbar)=\partial_a { \mathcal{F}}^{(N_f)} ({a}, {\bf m},\Lambda_{N_f}, \hbar)\Big |_{a=a(E,{\bf m},\Lambda_{N_f}, \hbar)}.
\ee

Here $ { \mathcal{F}}^{(N_f)} ({a}, {\bf m},\Lambda_{N_f}, \hbar)$ is the Nekrasov-Shatashvili free energy \cite{Nekrasov:2009rc}, and the quantum mirror map
$a$ is linked to $E$ through the Matone relation \cite{Matone:1995rx}:
\be
\label{matone}
{E=a^2-\dfrac{\Lambda_{N_f}}{4-N_f} \dfrac{\partial \mathcal{F}_{inst}^{(N_f)}(a;{\bf m};\Lambda_{N_f}, \hbar)}{\partial \Lambda_{N_f}}},
\ee 
where $\mathcal{F}_{inst}^{(N_f)}(a;{\bf m};\Lambda_{N_f}, \hbar)$ is the instanton part of $\mathcal{F}^{(N_f)}(a;{\bf m};\Lambda_{N_f}, \hbar)$.

\subsection{Quasinormal modes of Kerr black holes}
Choosing the Boyer-Lindquist coordinates, 
the metric of a four dimensional Kerr black hole in an  asymptotically flat spacetime is given by \cite{Detweiler:1980uk}
{
\ba
ds^2 &=&
-\left(1-\frac{2Mr}{\Sigma}\right)dt^2+\frac{\Sigma}{\Delta}dr^2-\frac{4\a Mr}{\Sigma}\sin^2\theta dt \,d\phi
+\Sigma d \theta^2\no\\
&&+\left(r^2+\a^2+\frac{2M\a^2r}{\Sigma}\sin^2\theta\right)\sin^2\theta \, d\phi^2, 
\ea}
where 
\be
\Sigma=r^2+\a^2\cos^2\theta, \quad \Delta = r^2-2M r+\a^2 \equiv(r-r_+)(r-r_-), 
\ee
with $M$ and $\alpha$ the mass and angular momentum of the black hole, respectively.
The singular points at $r_{\pm} \equiv M\pm\sqrt{M^{2}-\alpha^{2}}$ correspond to the Cauchy (plus sign) and event (minus sign) horizons.
If we write the scalar perturbations with a spin-$s$ in the Fourier space as
\be
\psi(t,r,\theta,\phi) = \frac{1}{2\pi} \int d\omega e^{-i \omega t} \sum_{l = |s|}^{\infty} \sum_{m=-l}^{l} e^{i m \phi} {}_{s}S_{\ell m}(\theta) R_{\ell m}(r),
\ee
the angular part (spin-weighted spheroidal harmonics) ${}_{s}S_{\ell m}(\theta)$ and the radial part $R_{\ell m}(r)$ satisfy the separated differential equations, the so-called Teukolsky equation. 
The angular part of the Teukolsky equation is given by
\begin{equation}
\begin{aligned}\label{angularT}
\biggl[ \frac{d}{dx}(1-x^2)\frac{d}{dx}+(cx)^2-2csx+{}_sA_{\ell m}+s-\frac{(m+sx)^2}{1-x^2} \biggr] {}_sS_{lm}(x)=0,
\end{aligned}
\end{equation}
where $x=\cos \theta$, $c=\alpha \omega$, and
 $\ell = 0, 1, 2\cdots$ with $|m|\leq \ell$. 
The eigenvalue ${}_sA_{\ell m}$ is determined by imposing the regularity condition at $x=\pm1$.
For the $c=0$ limit, the  spin-weighted spheroidal harmonics  ${}_sS_{lm}(x)$  reduces to the  spin-weighted spherical harmonics ${}_sY_{lm}$ and
its eigenvalue ${}_sA_{\ell m}$ becomes
\be{}_sA_{\ell m}(c=0)=\ell(\ell+1)-s(s+1). \ee

On the other hand, the radial part of the Teukolsky equation is 
\begin{equation}
\begin{aligned}\label{radialT}
\Delta(r) R''(r)+(s+1)\Delta'(r) R'(r)+V_T(r) R(r)=0,
\end{aligned}
\end{equation}
with $\Delta(r)=r^2-2Mr+\alpha^2$. Here, the potential is given by
\begin{equation}
\begin{aligned}
V_T(r)=\frac{K(r)^2-2is(r-M)K(r)}{\Delta(r)}-{}_sA_{\ell m}+4is\omega r+2\alpha m\omega-\alpha^2\omega^2,
\end{aligned}
\end{equation}
where $K(r)=(r^2+\alpha^2)\omega-\alpha m$.
To find the QNM, we have to impose the appropriate boundary conditions at the horizon and spatial infinity.
We choose the ingoing wave at the Cauchy horizon ($r=r_{+}$), whereas outgoing wave at the spatial infinity ($r=\infty$)
\begin{eqnarray}
    R(r\to r_{+}) &\sim& (r - r_{+})^{-s-i\sigma}, \\
    R(r \to \infty) &\sim& r^{-1-2s+2iM\omega}e^{i\omega r},
\end{eqnarray}
where 
\begin{equation}
    \sigma= \frac{2Mr_{+}\omega - \alpha m}{r_{+}-r_{-}}.
\end{equation}

For the angular part , after changing the variable $z=(1+x)/2$, and introducing $y(z)\equiv\sqrt{1-x^2}{}_sS_{lm}(x)/2$, \eqref{angularT} becomes
\begin{equation}
\begin{aligned}
\label{radial-2}
y''(z)+\left(\frac{1}{z^2(z-1)^2} \sum_{i=0}^4 A_i z^i\right)y(z)=0,
\end{aligned}
\end{equation}
also in the form of confluent Heun equation, and
 each coefficient is explicitly written as 
\begin{equation}
\begin{aligned}
    A_{0} &= \frac{1-(m-s)^{2}}{4}, \\
    A_{1} &= s(1-m)+(c+s)^{2}+{}_{s}A_{\ell m}, \\
    A_{2} &= -s-(c+s)(5c+s)- {}_{s}A_{\ell m}, \\
    A_{3} &= 4c(2c+s),\\
    A_{4} &= -4c^{2}.
\end{aligned}
\end{equation}

Likewise, for the radial part, defining $z=(r-r_-)/(r_+-r_-)$ and $y(z)\equiv\Delta(r)^{(s+1)/2} R(r)$,
we obtain the same form as \eqref{radial-2}  with different coefficients as follows:
{\small
\begin{equation}
\begin{aligned}
    A_{0} &= \frac{1}{4} \left[\frac{\left(\alpha  m+2 M \omega  \left(\sqrt{M^2-\alpha ^2}-M\right)\right)^2}{M^2-\alpha ^2}-\frac{2 i s \left(\alpha  m+2 M \omega   \left(\sqrt{M^2-\alpha ^2}-M\right)\right)}{\sqrt{M^2-\alpha ^2}}-s^2+1\right], \\
    A_{1} &=  \frac{i \alpha  m (s+2 i M \omega )-2 \omega  \left(2 M \omega  \left(\alpha ^2-2 M^2\right)+i \alpha ^2 s\right)}{\sqrt{M^2-\alpha ^2}}+s^{2}+s-\omega^{2}(8M^{2}-\alpha^{2}) {}_{s}A_{\ell m}, \\
    A_{2} &= -\omega^{2}  \left(5 \alpha ^2+12 M \sqrt{M^2-\alpha ^2}-12 M^2\right)- 6 i s \omega \sqrt{M^2-\alpha ^2} -s^{2}-s - {}_{s}A_{\ell m}, \\
    A_{3} &= -8\omega^{2}(M^{2}-\alpha^{2})+4\omega (i s+2 M \omega)   \sqrt{M^{2}-\alpha^{2}},\\
    A_{4} &= 4\omega^{2}(M^{2}-\alpha^{2}).
\end{aligned}
\end{equation}
}
Now that both the QNM side and the SW side satisfy the same confluent Heun equation form, we can find a dictionary by identifying the coefficients.

Explicitly, for the {\it angular} part, we find 
\begin{equation} 
\begin{aligned}
\label{angularDict}
\Lambda_3&=16c,\quad E=-{}_sA_{\ell m}-s(s+1)-c^2-\frac{1}{4}, \\
m_1&=-m, \quad m_2=m_3=-s.
\end{aligned}
\end{equation}

For the {\it radial} part, we have
\begin{equation} 
\begin{aligned}
\label{eq:id}
\Lambda_3&=-16\ri \omega \sqrt{M^2-\alpha^2}, \\
E&=-{}_sA_{\ell m}-s(s+1)+(8M^2-\alpha^2)\omega^2-\frac{1}{4}, \\
m_1&=s { -}2\ri M\omega,\qquad m_3=-s{ -}2\ri M\omega, \\
m_2&=\frac{\ri(-2M^2\omega+\alpha m)}{\sqrt{M^2-\alpha^2}}. 
\end{aligned}
\end{equation}
Especially when $\alpha=0$, it reproduces the identification in the Schwarzschild case written explicitly in  \cite{Aminov:2020yma} by exchanging $m_2 \leftrightarrow m_3$.

Then by substituting the QNM parameters into the quantization condition, we find that the angular part applies to the A-circle boundary condition, and similarly the radial part satisfies the B-circle condition as below,
\ba
&& \Pi_{B}^{(3)}\left(-{}_sA_{\ell m}-s(s+1)+(8M^2-\alpha^2)\omega^2-\frac{1}{4}  ,{\bf m},-16\ri \omega \sqrt{M^2-\alpha^2},1\right) =2\,\pi \left(n+{1\over 2}\right),\nonumber\\
&&{\bf m}=\biggl\{s{-}2\ri M\omega,\frac{\ri(-2M^2\omega+\alpha m)}{\sqrt{M^2-\alpha ^2}},-s{-}2\ri M\omega \biggr\} . \label{eq:BperiodKerr}
\ea
Through this equation, the frequency $\omega$ can be solved.

\section{Dictionary between Seiberg-Witten and Superradiance } \label{sec:3}
The method of separation of variables applied to the Klein-Gordon equation has been extensively evaluated in \cite{Detweiler:1980uk}, where the author presented a detailed analysis of the solutions obtained through this technique. In this section, we establish a novel connection by constructing a dictionary between the SW curve and the BH Superradiance, thereby bridging these two significant concepts.

For a massive scalar boson field $\psi$ around a BH, it obeys the
wave equation
\be\label{cmscalar}
\nabla^{a}\nabla_{\!a}\psi=\mu^2\psi\,,
\ee
with $\mu ={\cal M}G/\hbar c$ for a  particle of mass ${\cal M}$.
Equation \eqref{cmscalar} is also separable in the Kerr geometry. When we assume
\be
\psi=e^{-i\omega t+im\phi}S(\theta)R(r),
\ee
the separate equations are similar as before. For the angular part, we obtain:
\begin{align}\label{Seq}
\frac{1}{\sin\theta}\frac{d}{d \theta}\!
   \left[{\sin \theta}\frac{d S}{d \theta}\right]
   +\left[\alpha^2(\omega^2-\mu^2)\cos^2\theta-\frac{m^2}{\sin^2\theta}+{}_sA_{\ell m} \right]S=0\,,
\end{align}
and for the radial part:
\begin{align}\label{Req}
\Delta\frac{d}{d r}\!
  \left[{\Delta}\frac{d R}{d r}\right]
  +\left[\omega^2(r^2+\alpha^2)^2-4aMrm\omega+\alpha^2m^2-\Delta(\mu^2r^2+\alpha^2\omega^2+{}_sA_{\ell m})\right]R=0\,.
\end{align}
Or equivalently
\begin{equation}
\begin{aligned}
\Delta(r) R''(r)+\Delta'(r) R'(r)+\left(V_T(r)-\mu^2r^2\right) R(r)|_{s=0}=0,
\end{aligned}
\end{equation}
where we set the spin $s=0$ because we focus only on the scalar field here.
The quantity ${}_sA_{\ell m}$ is the separation constant, to be
found as an eigenvalue of \eqref{Seq}, and the eigenfunctions
$S(\theta)$ are the spheroidal harmonics. 
Changing the variables and following the same procedure in the previous section, we find that it is also in the form of confluent Heun equation \eqref{radial-2}, with
\begin{equation}
\begin{aligned}
    A_{0} &= \frac{1}{4} \left[\frac{\left(\alpha  m+2 M \omega  \left(\sqrt{M^2-\alpha ^2}-M\right)\right)^2}{M^2-\alpha ^2}+1\right], \\
    A_{1} &=  -2M^{2}(4 \omega^{2}-\mu^{2})-\alpha^{2}(\omega^{2}-\mu^{2}) \\
    &+\frac{2M}{\sqrt{M^{2}-\alpha^{2}}} \left( M^{2}(4\omega^{2}-\mu^{2})-\alpha^{2}(2\omega^{2}-\mu^{2}) - m \alpha \omega \right)-{}_{0}A_{\ell m}, \\
    A_{2} &= -5\alpha^2 (\omega^{2}-\mu^{2}) + 6 M(M-\sqrt{M^{2}-\alpha^{2}})(2\omega^{2}-\mu^{2})- {}_{0}A_{\ell m}, \\
    A_{3} &= -8(M^{2}-\alpha^{2})(\omega^{2}-\mu^{2})+4 M (2\omega^{2}-\mu^{2})   \sqrt{M^{2}-\alpha^{2}},\\
    A_{4} &= 4(M^{2}-\alpha^{2})(\omega^{2}-\mu^{2}).
\end{aligned}
\end{equation}
Comparing the potential of the wave equation in the gauge theory side, we obtain the dictionary for the massive scalar field.
For the radial part, we find
\begin{equation} 
\begin{aligned}
\Lambda_3&=-16 \ri \sqrt{\omega^2-\mu^2} \sqrt{M^2-\alpha^2}, \\
E&=-{}_0A_{\ell m}-(2M^2-\alpha^2)\mu^2+(8M^2-\alpha^2)\omega^2-\frac{1}{4}, \\
m_1&= -2\ri  M\omega,\qquad m_3=-\frac{ \ri M(2\omega^2-\mu^2)}{\sqrt{\omega^2-\mu^2}}, \\
m_2&=\frac{\ri(-2M^2\omega+\alpha m)}{\sqrt{M^2-\alpha^2}}. 
\end{aligned}
\label{eq:id}
\end{equation}
For the angular part, we have the same dictionary as the case of the massless scalar field \eqref{angularDict} with $c=\alpha\sqrt{\omega^{2}-\mu^{2}}$.
As in the previous section, the angular eigenvalue ${}_0A_{\ell m}$ can be expanded in $c$ as
\begin{equation}
    {}_0A_{\ell m} = \ell(\ell +1) + \sum_{k=1}^{\infty} f_{k} c^{2k}, 
\end{equation}
and we employ up to the $c^{6}$ order in our numerical calculations. 
Note that if we take $\mu=0$, the dictionary straightforwardly reduces to the identifications in the massless scalar case.

\section{Calculation by Instanton Counting} \label{sec:4}

As proposed in \cite{Aminov:2020yma}, 
the eigenvalues of the angular part are determined by the quantum A-period whereas the complex frequencies can be obtained by the quantum B-period, where each parameter is replaced according to the aforementioned dictionary. 
In the following, we discuss the relation between the quantum B-period and the boundary conditions we have to impose.

\subsection{Connection formula}
To compute the complex eigenfrequencies for the radial part of Teukolsky equation, we need to impose the boundary conditions at the horizon and spatial infinity.
In general, we obtain the following asymptotic behaviors of the field at the horizon and spatial infinity:
\begin{eqnarray}
    R(r\to r_{+}) &\sim& (r-r_{+})^{\pm i \sigma}, \label{eq:inout}\\
    R(r\to \infty) &\sim& r^{-1} r^{M(\mu^{2}-2\omega^{2})/q} e^{qr}, \label{eq:infinity}
\end{eqnarray}
where
\begin{equation}
    \sigma = \frac{2Mr_{+}\omega - \alpha m }{r_{+}-r_{-}} , \quad q = \pm\sqrt{\mu^{2}-\omega^{2}}.
\end{equation}
The sign of the exponent in (\ref{eq:inout}) corresponds to an outgoing and ingoing wave near the horizon.
Also, the sign of the real part of $q$ in (\ref{eq:infinity}) determines the asymptotic behavior of the wave as $r\to \infty$, corresponding to divergent or decaying, respectively.
The QNMs are ingoing at the horizon and purely outgoing $({\rm Re}[q]>0)$ at spatial infinity, whereas the so-called ``quasi-bound state'' solutions are ingoing at the horizon and decaying $({\rm Re}[q]<0)$ at spatial infinity. 

In terms of the field $y(z)$ solving the radial equation in Schr\"{o}dinger form, the above boundary conditions are rewritten as
\begin{eqnarray}
    y(z\to 1) &\sim& (z-1)^{\frac{1}{2}\pm  i \sigma}, \\
    y(z\to\infty) &\sim& \left(\Lambda_{3} z \right)^{\mp m_{3}} e^{\mp \frac{\Lambda_{3}z}{8} }. \label{eq:infinityY}
\end{eqnarray}
Note that the asymptotic behavior at spatial infinity is written in terms of the quantities in the gauge theory.
{According to the connection formula \cite{Bonelli:2022ten}, asymptotic expansions at different boundaries are related each other due to crossing symmetry.}
Now if we impose the ingoing wave boundary condition at the horizon, $y(z\to 1)\sim (z-1)^{\frac{1}{2}-i\sigma}$, and use the connection formula studied in \cite{Bonelli:2021uvf}, we can derive the asymptotic behavior at spatial infinity as
\begin{equation}
    y(z\to\infty) \sim C_{1}(\Lambda_{3}, a,{\bf m}) \left(\Lambda_{3} z \right)^{+ m_{3}} e^{+ \frac{\Lambda_{3}z}{8} } +C_{2}(\Lambda_{3}, a,{\bf m}) \left(\Lambda_{3} z \right)^{-m_{3}} e^{-\frac{\Lambda_{3}z}{8} },
\end{equation}
which is written as the linear combination of the asymptotic behaviors in (\ref{eq:infinityY}).
The explicit forms of $C_{1}$ and $C_{2}$ are shown in Appendix \ref{app:B}.
Demanding $C_{1}=0$ or $C_{2}=0$ is obviously corresponding to the choice of boundary conditions at spatial infinity, leading to the QNM or quasi-bound state, respectively.
In analogy with the discussion on the connection formula in \cite{Bonelli:2021uvf}, we find that the quantization condition for the QNM frequencies reads 
\ba
&& \Pi_{B}^{(3)}\left(-{}_0A_{\ell m}-(2M^2-\alpha^2)\mu^2+(8M^2-\alpha^2)\omega^2-\frac{1}{4} ,{\bf m},-16 \ri \sqrt{\omega^2-\mu^2} \sqrt{M^2-\alpha^2},1\right) \nonumber\\
&&=2\,\pi \left(n+{1\over 2} \right),\label{eq:BforQNM}\\
&&{\bf m}=\biggl\{-2\ri  M\omega,\frac{\ri(-2M^2\omega+\alpha m)}{\sqrt{M^2-\alpha ^2}},-\frac{ \ri M(2\omega^2-\mu^2)}{\sqrt{\omega^2-\mu^2}} \biggr\} .\nonumber
\ea
Additionally, we find that the quantization condition for the quasi-bound state frequencies is given by the same form as (\ref{eq:BforQNM}) with the sign of $m_{3}$ flipped, that is
\begin{equation}
    {\bf m}=\biggl\{-2\ri  M\omega,\frac{\ri(-2M^2\omega+\alpha m)}{\sqrt{M^2-\alpha ^2}},+\frac{ \ri M(2\omega^2-\mu^2)}{\sqrt{\omega^2-\mu^2}} \biggr\} . \label{BforQB}
\end{equation}
The details for the derivation of the quantization conditions are discussed in Appendix \ref{app:B}.

\subsection{Instanton counting}
{
With the above quantization conditions in hand, the frequency $\omega$ can be solved accordingly, making use of the relation \eqref{BtoF}, with the following instaton counting method:

The full SU(2) Nekrasov-Shatashvili free energy $ { \mathcal{F}}^{(N_f)} ({a}; {\bf m};\Lambda_{N_f}, \hbar)$ \cite{Nekrasov:2009rc} has contributions from its classical, one-loop and instanton components, with 
 this relation given explicitly as \cite{Aminov:2020yma}
\begin{equation}
\begin{aligned}
&\partial_a { \mathcal{F}}^{({N_f})} ({a}; {\bf m};\Lambda_{N_f}, \hbar) 
= -2\,a({4-N_f})\log{\left [\frac{\Lambda_{N_f} 2^{-{1\over (2-N_f/2)}}}{\hbar }\right ]}-\pi\hbar\\
&-2\,\ri\,\hbar\log{\left [
\dfrac{\Gamma({1+\frac{2\ri a}{\hbar})} } {\Gamma(1-\frac{2\ri a}{\hbar})}\right ]}
-\ri\,\hbar
\sum_{j=1}^{N_f}\log {\left [\dfrac{\Gamma ({\frac12+\frac{m_j-\ri a}{\hbar}})}
{\Gamma ({\frac12+\frac{m_j+\ri a}{\hbar}})}\right ]} +\dfrac{\partial { \mathcal{F}}_{\rm inst}^{(N_f)}({a};    {\bf m}  ; \Lambda_{N_f}, \hbar) }{\partial a}. 
\end{aligned}
\end{equation}

The SU(2) instanton part ${ \mathcal{F}}_{\rm inst}^{(N_f)}({a};    {\bf m}  ; \Lambda_{N_f}, \hbar)$ can be obtained by removing the U(1) contribution in the U(2)
 instanton part $F_{\rm inst}^{(N_f)}({a}; {\bf m} ;  \Lambda_{N_f},\hbar)$, which  is defined by 
\begin{equation}
\label{}
F_{\rm inst}^{(N_f)}({a}; {\bf m} ;  \Lambda_{N_f},\hbar) = -\hbar \, \lim_{\epsilon_2
  \rightarrow 0} \, \epsilon_2 \log Z^{(N_f)}({ \ri}{a},   {\bf m} ,  \hbar , \epsilon_2).
\end{equation}

The Nekrasov partition function $Z^{(N_f)}({ \ri}{a},   {\bf m} ,  \epsilon_1 , \epsilon_2)$ is exact in $\epsilon_i$, and can be written explicitly in terms of $\Lambda_{N_f}$ instanton expansion \cite{Alday:2009aq}, as a convergent series. 
In practice, we truncate this expansion to certain order (in this paper, $\Lambda_{N_f}^5$), and in general higher precision (depending on the convergence rate) will be obtained when we apply higher order expansions.
}

\subsection{Massless scalar fields in Schwarzschild and Kerr geometry}
The QNM frequencies for the massless scalar mode in the Schwarzschild and Kerr geometry were computed by imposing the quantum B-period in \cite{Aminov:2020yma}.
The explicit quantization condition is (\ref{eq:BperiodKerr}), which corresponds to (\ref{eq:BforQNM}) when taking the limit of $\mu\to 0$.
Hereafter we set $M=1$, so that $\omega$ and $\mu$ are measured in units of $M^{-1}$ and $\alpha$ in units of $M$.
We show the complex frequencies in table \ref{tab:tabSch} for the Schwarzschild geometry and table \ref{tab:tabKerr} for the Kerr geometry as the consistency check of our numerical setups.
The middle row shows the numerical values obtained from \cite{Berti2009quasinormal,Emanuele} and the right row shows our numerical results obtained from the SW/QNM correspondence.
\begin{table}[h!]
    \centering
    \caption{The complex frequencies $2\omega_{n}$ of the QNM in the Schwarzschild geometry.}
    \includegraphics[width=0.7\linewidth]{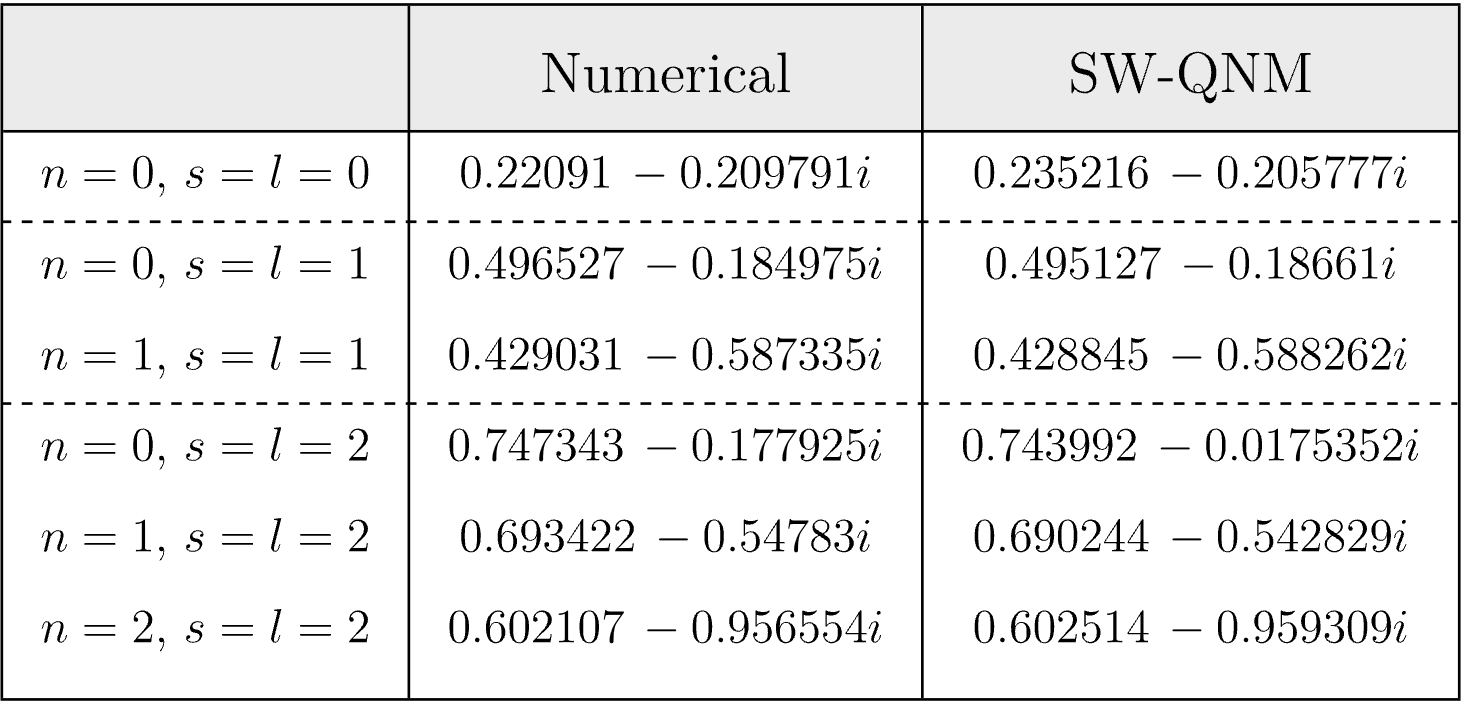}
    \label{tab:tabSch}
\end{table}
\begin{table}[h!]
    \centering
    \caption{The complex frequencies $\omega_{n}$ of the QNM in the Kerr geometry.}
    \includegraphics[width=\linewidth]{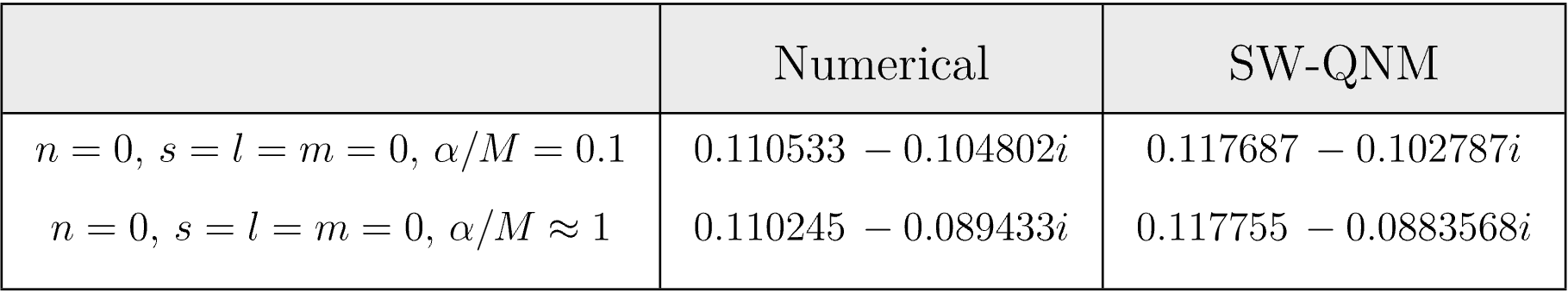}
    \label{tab:tabKerr}
\end{table}
In our actual calculations, we truncate the instanton counting series ${\cal{F}}^{(3)}_{\rm inst}$ at the fifth order in $\Lambda_{3}$.
We consider that the deviations between the numerical values and our results are due to numerical errors caused by omitting higher orders of $\Lambda_{3}$ in the instanton counting series.

\subsection{Massive scalar fields in Kerr geometry}
\begin{figure}[tbp]
    \centering
    \includegraphics[width=0.8\linewidth]{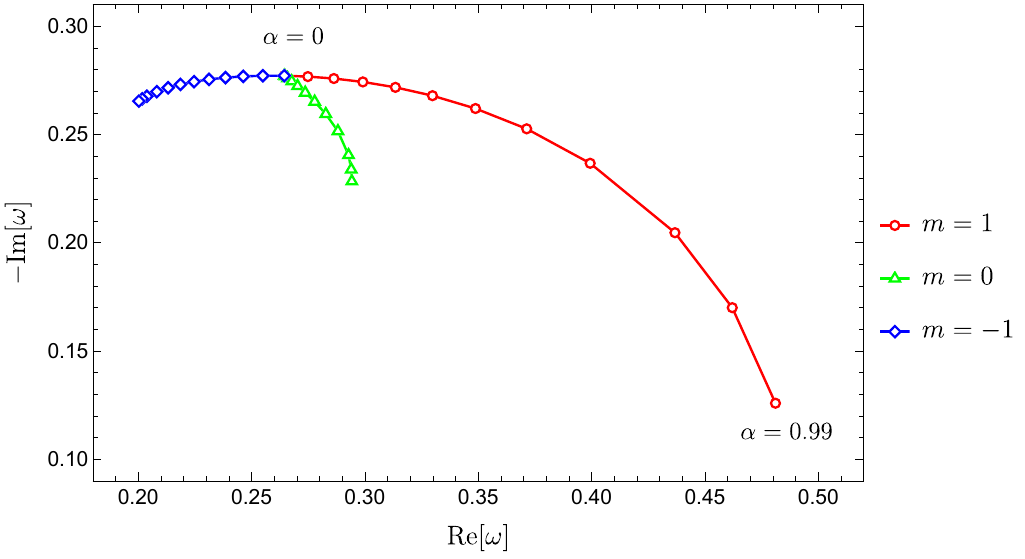}
    \caption{The QNM with $l=1$, $m=0, \pm1$, and $\mu=0.3$ as a function of the rotating parameter $\alpha$ in the complex frequency plane. The points denote each plot for from $\alpha=0$ to $\alpha=0.99$.}
    \label{fig:plotqnm}
\end{figure}

In this section, we investigate the complex eigenfrequencies for massive scalar fields.
As discussed in the previous section, the spectrum of massive scalar perturbations in the Kerr metric contains both stable QNM and modes of quasi-bound states depending on the boundary conditions at spatial infinity.
The quasi-bound states become unstable in the superradiant regime \cite{Dolan:2007mj}.
Firstly, we find that the complex frequencies for stable QNM can be obtained by imposing the quantum B-period (\ref{eq:BforQNM}).
Figure \ref{fig:plotqnm} shows the behaviors of QNM with $l=1$ and $\mu=0.3$ in the complex frequency plane as a function of the rotating parameter $\alpha$.
The plot qualitatively agrees with the results obtained by the continued fraction method (see figure 1 in \cite{Dolan:2007mj}).
The imaginary part of frequencies seems to have a numerical error in the order of ${\cal{O}}(10^{-1})$, which should be improved by considering higher orders of $\Lambda_{3}$.

Now let us confirm if our method can detect the superradiance instability for the quasi-bound state.
Here, we focus on the complex frequencies of the lowest quasi-bound state with $l=m=1$ and the fixed rotating parameter $\alpha$.
Table \ref{tab:omegatable} shows the complex frequencies of the quasi-bound state for $\alpha = 0.99$ near the superradiant regime with the continued fraction method and the SW/QNM correspondence.
\begin{table}[h!]
    \centering
    \caption{The complex frequencies $\omega/\mu$ of the quasi-bound state for $\alpha = 0.99$ and several values of $M\mu$ near the superradiant regime. The values of Continued fraction method are obtained following \cite{Dolan:2007mj}.}\vspace{0.5em}
    \includegraphics[width=0.8\linewidth]{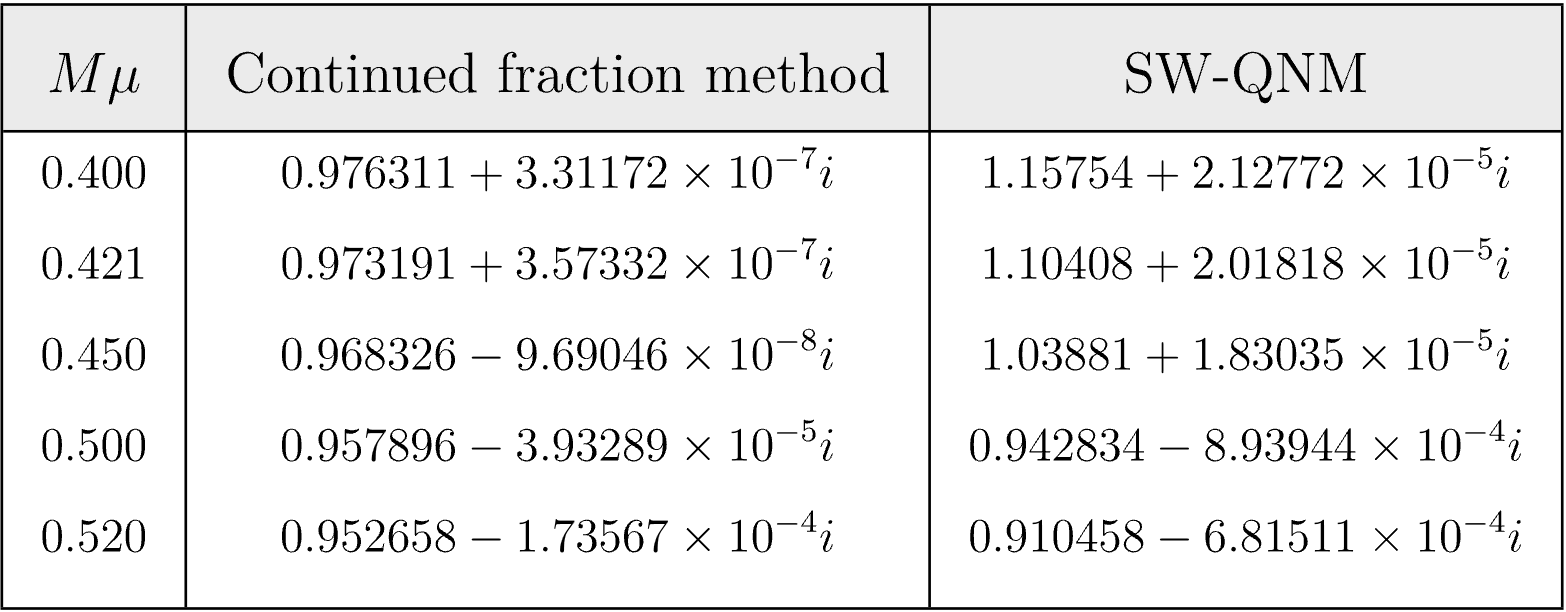}
    \label{tab:omegatable}
\end{table}
Compared to the known values, we find that the complex frequencies by the SW/QNM correspondence are similar, though have some numerical deviations, especially for the imaginary parts, caused by the numerical error (and also because the imaginary parts are much smaller than the real parts).
Nonetheless, the flip of the sign in the imaginary part of complex frequency around $\mu \approx 0.450$ are supposed to qualitatively imply the superradiant instability. It  clearly shows that the analytical SW/QNM method works for superradiance.

\section{Conclusions} \label{sec:5}
In our study, we propose the dictionary between the perturbative equations for massive scalar fields in the four-dimensional Kerr BH and the quantum SW curve by extending the original finding \cite{Aminov:2020yma}.
We find that the quantization conditions on the gauge theory lead to the complex frequencies for both the stable QNM and quasi-bound states by using the corresponding dictionaries derived from the connection formula.
The superradiant instability  occurs in the latter case and it is qualitatively implied by our calculations.
Hence,  our work establishes a profound connection between the SW theory and the phenomenon of BH superradiance. This connection is forged through their shared link to confluent Heun equations, which provide a mathematical bridge between the two seemingly disparate fields. By constructing a duality dictionary, we demonstrate that the frequencies of superradiance can be expressed analytically in terms of gauge parameters. This analytical relationship offers a unique perspective on superradiance, allowing us to gain deeper insights into its underlying physics.

In upcoming studies, we plan to establish stronger ties between gravitational atoms involving (scalar and vector) ultralight bosons and the SW curve. By doing so, we hope to not only refine our comprehension of these theoretical constructs but also develop novel techniques for detecting evidence of these phenomena through advanced GW observation analysis, thereby enriching our knowledge of fundamental physics.




\section*{{\it Acknowledgements.} }

We would like to thank Yang Lei, Hongfei Shu and Rui-Dong Zhu for very inspiring discussions. The work of XHG is supported in part by NSFC, China (Grant No. 12275166 and No. 12311540141). MM is supported by Shanghai Post-doctoral Excellence Program. KZ (Hong Zhang) is supported by a classified fund of Shanghai city.

\appendix

\section{Heun Equation and Its Confluent Generalizations}
The (general) Heun Equation (GHE) \cite{Heun:1888} reads:
\be\label{heun}
\frac{\mathrm{d}^2 w}{\mathrm{d} z^2} +\left({\g\over z}+{\d\over {z-1}}+{\e\over {z-a}}\right)\frac{\mathrm{d} w}{\mathrm{d} z} 
+\frac{\a\b z-q}{z(z-1)(z-a)}w=0\,,
\ee
with the constraint $\g+\d+\e=\a+\b+1$.

The Confluent Heun Equation (CHE) is
\be\label{heunC}
\frac{\mathrm{d}^2 w}{\mathrm{d} z^2} +\left({\hat\g\over z}+{\hat\d\over {z-1}}+{\hat\e}\right)\frac{\mathrm{d} w}{\mathrm{d} z} 
+\frac{\hat\a z-\hat q}{z(z-1)}w=0\,.
\ee

While the Biconfluent Heun Equation (BHE) is 
\be\label{heunb}
\frac{\mathrm{d}^2 w}{\mathrm{d} z^2} -\left({\tilde\g\over z}+{\tilde\d}+z\right)\frac{\mathrm{d} w}{\mathrm{d} z} 
+\frac{\tilde\a z-\tilde q}{z}w=0\,.
\ee

By the conflunce processes, we can obtain the latter two from GHE \cite{decarreau1978formes}. The transformation to CHE is straightforward, while for BHE,
 we can get this by coalescing three regular singularities. In details,
 we could change $z$ to $z/b$, so that \eqref {heun} becomes
\be\label{heunb1}
\frac{\mathrm{d}^2 w}{\mathrm{d} z^2} +\left(\frac{a b^2 \gamma +z \left(-a b (\gamma +\delta )-b (\gamma +\epsilon )\right)+z^2 (\gamma +\delta
   +\epsilon )}{z(z-b)(z-ab)}\right)\frac{\mathrm{d} w}{\mathrm{d} x} 
+\frac{\a\b z-b q}{z(z-b)(z-ab)}w=0\,.
\ee
Taking the limit  $b\to \infty$, we have
\be\label{heunb2}
\frac{\mathrm{d}^2 w}{\mathrm{d} z^2} +\left(\frac{  \gamma }{z}+ \left(- \frac{\gamma +\delta }{b}-\frac {\gamma +\epsilon }{ab}\right)+z\frac{ \gamma +\delta
   +\epsilon }{ab^2}\right)\frac{\mathrm{d} w}{\mathrm{d} x} 
+\frac{{\a\b\over ab^2} z-{q\over ab}}{z}w=0\,.
\ee
Then we obtain \eqref{heunb}, after identifying the corresponding coefficients, i.e., $\gamma=-\tilde\g$, $\d=-{ab\over{1-a}}(\tilde\d +b)$, $\e={ab\over{1-a}}(\tilde\d +ab)$, etc..

Those equations can further be rewritten into the normal form \cite{decarreau1978formes, cheb2004solutions} by redefining $y\to F(z) y$ with proper $F(z)$. For GHE, we have:
\be
\frac{\mathrm{d}^2 w}{\mathrm{d} z^2} +\left({A\over z}+{B\over {z-1}}+{C\over {z-a}}+{D\over z^2}+{E\over {(z-1)}^2}+{F\over {(z-a)}^2}\right)w=0\,.
\ee

For CHE, its normal form is given by 
\be
\label{CHEnormal}
\frac{\mathrm{d}^2 w}{\mathrm{d} z^2} +\left(A+{B\over z}+{C\over {z-1}}+{D\over z^2}+{E\over {(z-1)}^2}\right)w=0\,.
\ee

For BHE, the normal form is 
\be
\frac{\mathrm{d}^2 w}{\mathrm{d} z^2} +\left(-z^2+{B  z}+{C}+{D\over z}+{E\over {z^2}}\right)w=0\,.
\ee

\section{Quantization Condition for Quasi-bound States} \label{app:B}
Following \cite{Bonelli:2021uvf} we derive the quantization condition for the calculations of quasi-bound state frequencies.
By imposing the ingoing boundary condition at the horizon, the field $y(z)$ for the radial part is written as
\begin{equation}
        y(z\to\infty) \sim C_{1}(\Lambda, a,{\bf m}) \left(\Lambda z \right)^{+ m_{3}} e^{+ \frac{\Lambda z}{2} } +C_{2}(\Lambda, a,{\bf m}) \left(\Lambda z \right)^{-m_{3}} e^{-\frac{\Lambda z}{2} },
\end{equation}
where we follow the convention in \cite{Bonelli:2021uvf} and $\Lambda$ relates to our convention via $\Lambda= \Lambda_{3}/4$.
Additionally, we have to change $a \to -i a$ to match our convention,  which gives a different U(1) factor and a sign difference in the definition of the free energy $\cF$.
The coefficients $C_{1,2}$ are explicitly given by
\begin{eqnarray}
    &&\begin{split}
    C_{1}(\Lambda, a,{\bf m}) = \Lambda^{a} &M_{\alpha_{2+},\alpha_{+}} {\cal{A}}_{\alpha_{+}m_{0+}} 
    \frac{
    \mel{\Delta_{\alpha+},\Lambda_{0},m_{0+}}{V_{\alpha_{2}}(1)}{\Delta_{\alpha_{1}}}
    }
    {
    \mel{\Delta_{\alpha},\Lambda_{0},m_{0}}{V_{\alpha_{2+}}(1)}{\Delta_{\alpha_{1}}}
    } \\
    &+
    \Lambda^{-a} M_{\alpha_{2+},\alpha_{-}} {\cal{A}}_{\alpha_{-}m_{0+}} 
    \frac{
    \mel{\Delta_{\alpha-},\Lambda_{0},m_{0+}}{V_{\alpha_{2}}(1)}{\Delta_{\alpha_{1}}}
    }
    {
    \mel{\Delta_{\alpha},\Lambda_{0},m_{0}}{V_{\alpha_{2+}}(1)}{\Delta_{\alpha_{1}}}
    },
    \end{split} \\
    &&\begin{split}
    C_{2}(\Lambda, a,{\bf m}) = \Lambda^{a} &M_{\alpha_{2+},\alpha_{+}} {\cal{A}}_{\alpha_{+}m_{0-}} 
    \frac{
    \mel{\Delta_{\alpha+},\Lambda_{0},m_{0-}}{V_{\alpha_{2}}(1)}{\Delta_{\alpha_{1}}}
    }
    {
    \mel{\Delta_{\alpha},\Lambda_{0},m_{0}}{V_{\alpha_{2+}}(1)}{\Delta_{\alpha_{1}}}
    } \\
    &+
    \Lambda^{-a} M_{\alpha_{2+},\alpha_{-}} {\cal{A}}_{\alpha_{-}m_{0-}} 
    \frac{
    \mel{\Delta_{\alpha-},\Lambda_{0},m_{0-}}{V_{\alpha_{2}}(1)}{\Delta_{\alpha_{1}}}
    }
    {
    \mel{\Delta_{\alpha},\Lambda_{0},m_{0}}{V_{\alpha_{2+}}(1)}{\Delta_{\alpha_{1}}}
    }.
    \end{split}
\end{eqnarray}
Here, we omit to write down the definitions of each factor because we use the same convention as in \cite{Bonelli:2021uvf}.
The condition of $C_{1}=0$ reads
\begin{equation}
    1+ \Lambda^{-2a} 
    \frac{
    M_{\alpha_{2+},\alpha_{-}} {\cal{A}}_{\alpha_{-}m_{0+}} \mel{\Delta_{\alpha-},\Lambda_{0},m_{0+}}{V_{\alpha_{2}}(1)}{\Delta_{\alpha_{1}}}
    }{
    M_{\alpha_{2+},\alpha_{+}} {\cal{A}}_{\alpha_{+}m_{0+}} \mel{\Delta_{\alpha+},\Lambda_{0},m_{0+}}{V_{\alpha_{2}}(1)}{\Delta_{\alpha_{1}}}
    } =0. \label{eq:C1cond}
\end{equation}
In the Nekrasov-Shatashvili  limit, 
\begin{align}
    &\frac{
    \mel{\Delta_{\alpha-},\Lambda_{0},m_{0+}}{V_{\alpha_{2}}(1)}{\Delta_{\alpha_{1}}}
    }{
    \mel{\Delta_{\alpha+},\Lambda_{0},m_{0+}}{V_{\alpha_{2}}(1)}{\Delta_{\alpha_{1}}}
    }
    =
    \frac{
    Z^{(3)}(\Lambda,a+\frac{\epsilon_{2}}{2}, m_{1}, m_{2}, m_{3}+ \frac{\epsilon_{2}}{2})
    }{
    Z^{(3)}(\Lambda,a-\frac{\epsilon_{2}}{2}, m_{1}, m_{2}, m_{3}+ \frac{\epsilon_{2}}{2})
    } \nonumber\\
    &= \exp \frac{1}{\epsilon_{1}\epsilon_{2}} \left[ \cF_{\rm inst}^{(3)} \left( \Lambda,a+\frac{\epsilon_{2}}{2}, m_{1}, m_{2}, m_{3}+ \frac{\epsilon_{2}}{2}\right) - \cF_{\rm inst}^{(3)} \left( \Lambda,a-\frac{\epsilon_{2}}{2}, m_{1}, m_{2}, m_{3}+ \frac{\epsilon_{2}}{2}\right) \right] \nonumber\\
    &\to \exp \frac{\partial_{a} \cF_{\rm inst}^{(3)}(\Lambda,a,m_{1},m_{2},m_{3})}{\epsilon_{1}},
\end{align}
and
\begin{eqnarray}
    \frac{
    M_{\alpha_{2+},\alpha_{-}} {\cal{A}}_{\alpha_{-}m_{0+}} 
    }{
    M_{\alpha_{2+},\alpha_{+}} {\cal{A}}_{\alpha_{+}m_{0+}} 
    } &= &
    \frac{
    \gmf{\frac{2a}{\eone}} \gmf{1+\frac{2a}{\eone}} \gmf{\frac{1}{2}+\frac{a_{2}+a_{1}-a}{\eone}} \gmf{\frac{1}{2}+\frac{a_{2}-a_{1}-a}{\eone}} \gmf{\frac{1}{2}+ \frac{m_{3}-a}{\eone}}
    }
    {
    \gmf{-\frac{2a}{\eone}} \gmf{1-\frac{2a}{\eone}} \gmf{\frac{1}{2}+\frac{a_{2}+a_{1}+a}{\eone}} \gmf{\frac{1}{2}+\frac{a_{2}-a_{1}+a}{\eone}} \gmf{\frac{1}{2}+ \frac{m_{3}+a}{\eone}}
    } \nonumber\\
    &=& 
    \frac{
    \gmf{\frac{2a}{\eone}} \gmf{1+\frac{2a}{\eone}} 
    }
    {
    \gmf{-\frac{2a}{\eone}} \gmf{1-\frac{2a}{\eone}} 
    } \prod_{i=1}^{3} \frac{\gmf{\frac{1}{2}+\frac{m_{i}-a}{\eone}}}{\gmf{\frac{1}{2}+\frac{m_{i}+a}{\eone}}} \no \\
    &=&
    e^{-i\pi} \left(  \frac{
    \gmf{1+\frac{2a}{\eone}} 
    }
    {
    \gmf{1-\frac{2a}{\eone}} 
    } \right)^{2} \prod_{i=1}^{3} \frac{\gmf{\frac{1}{2}+\frac{m_{i}-a}{\eone}}}{\gmf{\frac{1}{2}+\frac{m_{i}+a}{\eone}}} \no \\
    &=& \exp \left[ -i\pi + 2 \log \frac{\gmf{1+ \frac{2a}{\eone}}}{\gmf{1- \frac{2a}{\eone}}} + \sum_{i=1}^{3} \log \frac{
    \gmf{\frac{1}{2}+\frac{m_{i}-a}{\eone}}
    }
    {
    \gmf{\frac{1}{2} + \frac{m_{i}+a}{\eone}}
    }\right],
\end{eqnarray}
where we use the relations of parameters between CFT and gauge theory
\begin{equation}
    m_{1} = a_{1}+a_{2}, \quad m_{2} = -a_{1} +a_{2}.
\end{equation}
Setting $\eone =1$, the second term of (\ref{eq:C1cond}) becomes
\begin{equation}
\begin{split}
    \exp \bigg[ 
    -2a \log\Lambda -i\pi + 2 \log \frac{\gmf{1+ \frac{2a}{\eone}}}{\gmf{1- \frac{2a}{\eone}}} &+ \sum_{i=1}^{3} \log \frac{
    \gmf{\frac{1}{2}+\frac{m_{i}-a}{\eone}}
    }
    {
    \gmf{\frac{1}{2} + \frac{m_{i}+a}{\eone}}
    }
    + \partial_{a} \cF_{\rm inst}^{(3)}
    \bigg] \\
    &= \exp \left[ \partial_{a}\cF^{(3)}(\Lambda, a, {\bf m}, \hbar=1)  \right].
\end{split}
\end{equation}
Therefore, we obtain the quantization condition for the QNM:
\begin{equation}
    \partial_{a}\cF^{(3)} (\Lambda,a,{\bf m }, 1) = i(2n+1) \pi, 
\end{equation}
where $n$ is an integer.

In a similar way, the condition of $C_{2}=0$ reads
\begin{equation}
    1+ \Lambda^{-2a} 
    \frac{
    M_{\alpha_{2+},\alpha_{-}} {\cal{A}}_{\alpha_{-}m_{0-}} \mel{\Delta_{\alpha-},\Lambda_{0},m_{0-}}{V_{\alpha_{2}}(1)}{\Delta_{\alpha_{1}}}
    }{
    M_{\alpha_{2+},\alpha_{+}} {\cal{A}}_{\alpha_{+}m_{0-}} \mel{\Delta_{\alpha+},\Lambda_{0},m_{0-}}{V_{\alpha_{2}}(1)}{\Delta_{\alpha_{1}}}
    } =0. \label{eq:C1cond}
\end{equation}
In the NS limit,
\begin{align}
    &\frac{
    \mel{\Delta_{\alpha-},\Lambda_{0},m_{0-}}{V_{\alpha_{2}}(1)}{\Delta_{\alpha_{1}}}
    }{
    \mel{\Delta_{\alpha+},\Lambda_{0},m_{0-}}{V_{\alpha_{2}}(1)}{\Delta_{\alpha_{1}}}
    }
    =
    \frac{
    Z^{(3)}(\Lambda,a+\frac{\epsilon_{2}}{2}, m_{1}, m_{2}, m_{3}- \frac{\epsilon_{2}}{2})
    }{
    Z^{(3)}(\Lambda,a-\frac{\epsilon_{2}}{2}, m_{1}, m_{2}, m_{3}- \frac{\epsilon_{2}}{2})
    } \nonumber\\
    &= \exp \frac{1}{\epsilon_{1}\epsilon_{2}} \left[ \cF_{\rm inst}^{(3)} \left( \Lambda,a+\frac{\epsilon_{2}}{2}, m_{1}, m_{2}, m_{3}- \frac{\epsilon_{2}}{2}\right) - \cF_{\rm inst}^{(3)} \left( \Lambda,a-\frac{\epsilon_{2}}{2}, m_{1}, m_{2}, m_{3}- \frac{\epsilon_{2}}{2}\right) \right] \nonumber\\
    &\to \exp \frac{\partial_{a} \cF_{\rm inst}^{(3)}(\Lambda,a,m_{1},m_{2},m_{3})}{\epsilon_{1}},
\end{align}
and
\begin{eqnarray}
    \frac{
    M_{\alpha_{2+},\alpha_{-}} {\cal{A}}_{\alpha_{-}m_{0-}} 
    }{
    M_{\alpha_{2+},\alpha_{+}} {\cal{A}}_{\alpha_{+}m_{0-}} 
    } &= &
    \frac{
    \gmf{\frac{2a}{\eone}} \gmf{1+\frac{2a}{\eone}} \gmf{\frac{1}{2}+\frac{a_{2}+a_{1}-a}{\eone}} \gmf{\frac{1}{2}+\frac{a_{2}-a_{1}-a}{\eone}} \gmf{\frac{1}{2}+ \frac{-m_{3}-a}{\eone}}
    }
    {
    \gmf{-\frac{2a}{\eone}} \gmf{1-\frac{2a}{\eone}} \gmf{\frac{1}{2}+\frac{a_{2}+a_{1}+a}{\eone}} \gmf{\frac{1}{2}+\frac{a_{2}-a_{1}+a}{\eone}} \gmf{\frac{1}{2}+ \frac{-m_{3}+a}{\eone}}
    } \nonumber\\
    &=& 
    \frac{
    \gmf{\frac{2a}{\eone}} \gmf{1+\frac{2a}{\eone}} 
    }
    {
    \gmf{-\frac{2a}{\eone}} \gmf{1-\frac{2a}{\eone}} 
    } \prod_{i=1}^{3} \frac{\gmf{\frac{1}{2}+\frac{\tilde{m}_{i}-a}{\eone}}}{\gmf{\frac{1}{2}+\frac{\tilde{m}_{i}+a}{\eone}}} \no \\
    &=&
    e^{-i\pi} \left(  \frac{
    \gmf{1+\frac{2a}{\eone}} 
    }
    {
    \gmf{1-\frac{2a}{\eone}} 
    } \right)^{2} \prod_{i=1}^{3} \frac{\gmf{\frac{1}{2}+\frac{\tilde{m}_{i}-a}{\eone}}}{\gmf{\frac{1}{2}+\frac{\tilde{m}_{i}+a}{\eone}}} \no \\
    &=& \exp \left[ -i\pi + 2 \log \frac{\gmf{1+ \frac{2a}{\eone}}}{\gmf{1- \frac{2a}{\eone}}} + \sum_{i=1}^{3} \log \frac{
    \gmf{\frac{1}{2}+\frac{\tilde{m}_{i}-a}{\eone}}
    }
    {
    \gmf{\frac{1}{2} + \frac{\tilde{m}_{i}+a}{\eone}}
    }\right],
\end{eqnarray}
where we define 
\begin{equation}
    \tilde{m}_{1} = m_{1}, \quad \tilde{m}_{2}=m_{2},\quad \tilde{m}_{3} = - m_{3}.
\end{equation}
Thus, the quantization condition for the quasi-bound states is 
\begin{equation}
    \partial_{a}\cF^{(3)} (\Lambda,a,\tilde{\bf m}, 1) = i(2n+1) \pi.
\end{equation}
Consequently, in order to compute the complex frequencies of quasi-bound states, we have to impose the quantum B-period with the dictionary in which the sign of $m_{3}$ is flipped.

\bibliographystyle{JHEP}
\bibliography{SW-QNM}

\end{document}